\documentclass[12pt,longbibliography,aps,groupedaddress,showpacs,nofootinbib,prd,reprint]{revtex4-2}
\usepackage{amsmath,amssymb,graphicx,amsfonts}
\usepackage{mathrsfs}
\usepackage{xcolor}
\usepackage[shortlabels]{enumitem}

\def\<#1>{\mathinner{\langle#1\rangle}}

\usepackage{hyperref}
\usepackage{setspace}

\begin{document}
\setlength{\parskip}{1pt}
\title{Gravitational-wave glitches: resonant islands and frequency jumps in non-integrable extreme-mass-ratio inspirals}

\author{Kyriakos Destounis}
\email{kyriakos.destounis@uni-tuebingen.de}
\affiliation{Theoretical Astrophysics, IAAT, University of T{\"u}bingen, 72076 T{\"u}bingen, Germany}

\author{Kostas D. Kokkotas}
\affiliation{Theoretical Astrophysics, IAAT, University of T{\"u}bingen, 72076 T{\"u}bingen, Germany}
 
\begin{abstract}
\noindent{The detection of gravitational waves from extreme-mass-ratio inspirals with upcoming space-borne detectors will allow for unprecedented tests of general relativity in the strong-field regime. Aside from assessing whether black holes are unequivocally described by the Kerr metric, such detections may place constraints on the degree of spacetime symmetry. In particular, depending on exactly how a hypothetical departure from the Kerr metric manifests, the Carter symmetry, which implies the integrability of the geodesic equations, may be broken. Here, we examine the gravitational waveforms associated with non-integrable extreme-mass-ratio inspirals involving a small-mass companion and a supermassive compact object of general relativity, namely the Manko-Novikov spacetime. We show that the waveforms displays sudden frequency jumps, when the companion transverses resonant islands. These findings demonstrate that such abrupt manifestations in the gravitational-wave frequencies are generic, have a genuine astrophysical origin and function as a distinctive signature of chaotic phenomena in extreme-mass-ratio binaries.}
\end{abstract}

\maketitle

\section{Introduction}

The direct detection of gravitational waves (GWs) by ground-based detectors, such as the Laser Interferometer Gravitational-wave Observatory (LIGO) and Virgo \cite{LIGOScientific:2018mvr,Abbott:2020niy}, has unlocked a new realm into the exploration of coalescing compact objects in the strong-field regime. The successors to ground-based detectors, the Laser Interferometer Space Antenna (LISA) and Taiji \cite{lisafreq,taiji}, will redefine our understanding of GW astronomy by probing the low-frequency regime of binary mergers \cite{Berti:2019xgr}. One of their primary targets are extreme-mass-ratio inspirals (EMRIs).

These binaries involve a supermassive compact object (the primary) and a small-mass companion (the secondary). The fact that one of the binary components has, essentially, negligible mass compared to the other, implies that the secondary object transverses the gravitational field generated by the primary following geodesic motion, at zeroth order. Including higher-order self-force corrections to the equations of motion of the secondary, obtained from black-hole (BH) perturbation theory \cite{Poisson:2011nh,Barack:2018yvs}, leads to an adiabatic drift between geodesics due to radiation reaction. Taking into consideration that EMRIs evolve slowly and complete $\sim 10^5$ cycles in the strong-field regime means that they will provide unmatched information about the governing theory of gravity \cite{ryan95,bert05,glamp06,schutz09,Barack:2006pq,Barausse:2020rsu}. 

While a plethora of BH solutions currently exist in general relativity (GR), astrophysical BHs are expected to be described by the Kerr metric \cite{Kerr:1963ud,rob75}. The governing equations of an object of negligible, but conserved, rest mass orbiting around a Kerr BH are integrable, due the stationarity, axisymmetry and the existence of a hidden symmetry; the Carter constant \cite{carter68}. Since integrability implies the absence of chaotic phenomena \cite{wilkins72,cont02}, an EMRI involving a Kerr primary (or any other compact object with integrable geodesics) will produce a waveform with slowly drifting fundamental frequencies and harmonics. If, however, the primary's geometry is significantly distorted due to the presence of accretion disks \cite{rez07,sem13} (though their influence is expected to be negligible \cite{Barausse:2014tra}), or GR does not provide an exact description of geometries surrounding compact objects, then the Kerr metric may not adequately describe supermassive BHs. Including slight, or otherwise, modifications to the Kerr description provides a reasonable framework to deal with the aforementioned subtleties with a drawback; integrability may be broken (though see \cite{joh13,papkok18,Moore:2017lxy}).

When a non-integrable perturbation is introduced into the Hamiltonian, the resulting system may exhibit regions where chaotic phenomena arise, in accord with the Kolmogorov-Arnold-Moser (KAM) and Poincar\'e-Birkhoff theorems \cite{moser62,arnold63,poin12,birk13}. Such ergodic regions may either surround resonances \cite{arnold89} or form close to the plunge \cite{Gair:2007kr}. In EMRIs, though, chaotic motion itself is not expected to be a prominent effect, therefore one must search for alternative generic features that clearly distinguish the existence of chaos. It turns out that a solid indicator of non-integrability is the passage through resonant islands \cite{apo09,luk,cont11,Lukes-Gerakopoulos:2014dpa,Zelenka:2017aqn,dest20}. 

A resonance occurs when two or more characteristic frequencies of a system match in integer-number ratios. When this is the case, the resonant condition is fulfilled \cite{cont02} and the orbit becomes periodic. The resonant orbit no longer samples the available phase space, in contrast with generic orbits which cover it densely. Although resonances generically occur in integrable systems, such as Kerr, non-integrable Kerr-like compact objects (henceforth referred to as `non-Kerr' objects) possess regimes of non-zero phase-space volume called resonant islands (`Birkhoff islands'), within which geodesics share the same rational ratio of orbital frequencies \cite{poin12,birk13}. 

Resonant orbits are ubiquitous in EMRIs and pose quite a challenge in data analysis, by virtue of their topological structure which differs significantly from that of generic orbits. When the secondary object crosses a resonance, the flux of GWs can be diminished or enhanced, leading to a shift in its trajectory and the phase of the resulting waveform \cite{Flanagan:2010cd,Flanagan:2012kg,Berry:2016bit}. Inspiraling objects may spend, roughly, a few hours \cite{Brink:2013nna,Brink:2015roa} or a few dozens of orbital cycles `near' resonance \cite{Ruangsri:2013hra} (the resonant condition is satisfied within an arbitrarily short range) for supermassive BHs with masses $\sim 10^6 M_\odot$, though the observation of a very long-lived (sustained) resonance is highly unlike, if not practically impossible \cite{vandeMeent:2013sza}.

Resonances surrounded by Birkhoff islands of non-Kerr objects, on the other hand, lead to the entrapment of small-mass companions in `perfect' resonance (the resonant condition is precisely satisfied) for a time comparable to the width of the island, and should introduce even greater difficulties in parameter estimation. At the orbital level, occupancy in a resonant island translates to a dynamical plateau of the ratio of radial and polar frequencies \cite{cont02} which can last for up to a week or a few hundred orbital cycles \cite{luk,Lukes-Gerakopoulos:2014dpa}. Surprisingly, even sustained resonances are likely to occur in non-Kerr EMRIs, where the trapping time of the companion inside a resonant island is, roughly, comparable to the inspiral timescale \cite{Lukes-Gerakopoulos:2021ybx}.

Recently, it was shown that the appearance of a dynamical plateau in the orbital frequencies of a non-Kerr EMRI induces a substantial `glitch' to the frequency evolution of a GW detected by LISA \cite{Destounis:2021mqv}. These abrupt frequency jumps can be understood as direct imprints of chaotic phenomena in EMRIs, which in turn may either designate the violation of the `no-hair' theorem \cite{luk,Destounis:2021mqv}, the existence of an $N>2$-body inspiral \cite{Barausse:2006vt,AmaroSeoane:2011id} or the presence of spin in the secondary inspiraling object \cite{kiu04,Zelenka:2019uke,Zelenka:2019nyp}. 

Here, we consider an EMRI consisting of a small-mass secondary object inspiraling into a non-Kerr compact object described by a particular subclass of the Manko-Novikov spacetime \cite{Manko92} which is an exact solution to the vacuum Einstein equations and includes a generic deviation to the quadrupole mass moment. Although the orbital dynamics of such bumpy spacetime has been extensively analyzed \cite{Gair:2007kr,luk}, we show that the passage through a resonant island corresponds to a sudden jump in the orbital elements of the inspiral. This abrupt phenomenon manifests itself into the gravitational waveform as a frequency glitch, in an identical manner to those found in \cite{Destounis:2021mqv}. Hence, we demonstrate that GW glitches are generic features of non-integrable EMRIs, are intrinsically linked to the fundamental spacetime symmetries of the system and signify the presence of chaotic phenomena during the inspiral.

\section{The Manko-Novikov spacetime}

The goal of this study is to examine the universality of GW glitches in non-integrable EMRIs, such as those found recently in \cite{Destounis:2021mqv}. To achieve this, we utilize a metric, namely the Manko-Novikov (MN) spacetime \cite{Manko92}, which is an exact solution to the vacuum Einstein field equations and generalizes the Kerr-Newman metric into a multiparametric family. Although this solution depends, in general, on an infinite amount of deformation parameters related to the multipole moments of the gravitational field, we will adopt a particular subclass \cite{Gair:2007kr,luk} which admits one extra parameter, besides the mass $M$ and spin $S$, which controls the integrability of the geodesic equations. This new deformation parameter, $q$, is a dimensionless number that measures the deviation of the mass quadrupole moment of the gravitational field from that of the corresponding Kerr metric. The first three, non-vanishing, mass and current multipole moments are, then, characterized by $M,\,S,\,q$ as \cite{Gair:2007kr,luk}
\begin{align} 
\label{moments}
M_0 &= M,\,\,\,	S_1 = \chi M^2,\,\,\, M_2 = -M^3 \left(\chi^2+q\right), 
\end{align}
where $\chi\equiv S/M^2$ is the dimensionless spin parameter, while higher order moments are characterized by higher order polynomials of $q$. A general stationary, axisymmetric, vacuum spacetime can be written in the Weyl-Papapetrou coordinates $(t,\rho,z,\phi)$ as \cite{Gair:2007kr,luk}
\begin{equation}
\label{line_element}
ds^2 = -f(dt-\omega d\phi)^2+\frac{1}{f} \left[e^{2\gamma} (d\rho^2+dz^2)+\rho^2 d\phi^2 \right].
\end{equation}
For the MN spacetime, $f,~\omega,~\gamma$ are functions of the prolate spheroidal coordinates $(x,y)$ with the following form:
\begin{align}
f &= e^{2 \psi}\frac{A}{B}, \label{ffunc} \\
\omega &= 2 k e^{-2 \psi}\frac{C}{A}-4 k \frac{\alpha}{1-\alpha^2}, \\
e^{2 \gamma} &= e^{2 \gamma^\prime}\frac{A}{(x^2-1)(1-\alpha^2)^2},
\label{fexpgam} \\
A &= (x^2-1)(1+a~b)^2-(1-y^2)(b-a)^2,\label{fA} \\
B &= [(x+1)+(x-1)a~b]^2+[(1+y)a+(1-y)b]^2,\label{fB} \\
C &= (x^2-1)(1+a~b)[(b-a)-y(a+b)] \nonumber \\
&+ (1-y^2)(b-a)[(1+a~b)+x(1-a~b)], \\
\psi &= \beta \frac{P_2}{R^3}, \label{fC}\\
\gamma^\prime &= \ln{\sqrt{\frac{x^2-1}{x^2-y^2}}}+\frac{3\beta^2}{2 R^6}
(P_3^2-P_2^2) \nonumber \\ &+ \beta \left(-2+\displaystyle{\sum_{\ell=0}^2}
\frac{x-y+(-1)^{2-\ell}(x+y)}{R^{\ell+1}}P_\ell\right), \label{fgampr}
\end{align}
\begin{align}
a &= -\alpha \exp {\left[-2\beta\left(-1+\displaystyle{\sum_{\ell=0}^2}
		\frac{(x-y)P_\ell}{R^{\ell+1}}\right)\right]}, \label{fa}\\
b &= \alpha \exp {\left[2\beta\left(1+\displaystyle{\sum_{\ell=0}^2}
	\frac{(-1)^{3-\ell}(x+y)P_\ell}{R^{\ell+1}}\right)\right]}, \label{fb}
\end{align}
where $R=\sqrt{x^2+y^2-1}$ and $P_\ell=P_\ell(xy/R)$ are the Legendre polynomials of order $l$ given by
\begin{equation}
P_\ell(w)=\frac{1}{2^\ell \ell!}
\left(\frac{d}{dz}\right)^\ell(w^2-1)^\ell.
\label{fLeg}
\end{equation}
The line element \eqref{line_element} is related to the parameters $\alpha,\,k,\,\beta$ which are defined as
\begin{equation}
\alpha=\frac{-1+\sqrt{1-\chi^2}}{\chi},\,\,\,
k=M\frac{1-\alpha^2}{1+\alpha^2},\,\,\,
\beta=q \frac{M^3}{k^3}.
\end{equation}
Consequently, choosing $M,\,\chi$ and $q$ completely defines the MN metric. The relation between prolate spheroidal coordinates $(x,y)$ and quasi-cylindrical coordinates $(\rho,z)$ is given by \cite{luk,Bambi:2013eb}
\begin{equation}
\rho=k \sqrt{(x^2-1)(1-y^2)},\quad z=k x y,
\end{equation}
while the inverse transformation reads
\begin{align}
x&=\frac{1}{2k}\left(\sqrt{\rho^2+\left(z+k\right)^2}+\sqrt{\rho^2+\left(z-k\right)^2}\right),\\
y&=\frac{1}{2k}\left(\sqrt{\rho^2+\left(z+k\right)^2}-\sqrt{\rho^2+\left(z-k\right)^2}\right).
\end{align}
In turn, the relation between $(x,y)$, $(\rho,z)$ and the standard Boyer-Lindquist coordinates $(r,\theta)$ is given by
\begin{equation}
r=kx+M,\quad\theta=\cos^{-1}y
\end{equation}
and 
\begin{align} \label{BoyLindr}
	r/M&=1+\sqrt{-\sigma_2+\sqrt{\sigma_2^{~2}+\sigma_1(z/M)^2}},\\
	\label{BoyLindtheta}
	\theta&=\cos^{-1} \left( {\frac{z/M}{(r/M)-1}} \right),
\end{align}
where
\begin{equation*}
\sigma_1=\chi^2-1,\quad\sigma_2=\frac{\sigma_1-(\rho/M)^2-(z/M)^2}{2}.
\end{equation*}

The MN spacetime, having a multipolar structure different from that of Kerr, is not a BH solution, in accordance with the no-hair theorem. In fact, it lacks a compact event horizon and contains closed time-like curves, where $g_{\phi\phi}<0$,  exterior to its horizon which is broken in the equatorial plane by a circular line singularity. For non-zero spin parameters, the metric contains an ergoregion, determined by $g_{tt}=0$, which generally admits a multiple lobed structure \cite{Gair:2007kr}. For $q=0$, the Kerr metric is recovered, while a choice of $q>0\,(q<0)$ represents an oblate (prolate) perturbation of the Kerr metric.

Regardless of its pathological nature, the MN metric is a prime candidate to describe the gravitational field of a stationary axisymmetric compact object with a well-defined multipolar structure which deviates from that of Kerr. In particular, it does not admit a Carter constant, as opposed to Kerr, and thus provides a proper testing ground for the no-hair theorem, and its consequences on the waveforms of non-integrable inspirals. In what follows, we will restrict our numerical evolutions into regions where the near-horizon pathological features of the MN spacetime are not present.

\section{Short timescale EMRI modeling: geodesic motion}\label{sect:geod}
Due to GW emission, the orbital dynamics that governs an EMRI is not purely geodesic. Nevertheless, at zeroth-order and for short timescales, we can neglect the structure of the secondary and adopt a point-particle approximation, with its motion on the gravitational field created by the primary being described by the geodesic equation
\begin{equation}\label{geodesic}
\frac{d^2 x^\kappa}{d\tau^2}+\Gamma^\kappa_{\lambda\nu}\frac{dx^\lambda}{d\tau}\frac{dx^\nu}{d\tau}=0,
\end{equation}
where $\Gamma^\alpha_{\lambda\nu}$ are the Christoffel symbols, $x^\kappa$ the $4$-position vector and $\tau$ the proper time. Equivalently, one could operate in the Lagrange formulation to describe geodesic motion. The equations of motion for a secondary object with rest mass $\mu$ can be derived, via the Euler-Lagrange equations, by considering a Lagrangian 
\begin{equation}
\mathcal{L}=\frac{1}{2}\mu\, g_{\lambda\nu}\dot{x}^\lambda\dot{x}^\nu,
\end{equation}
where $g_{\lambda\nu}$ are the metric tensor components of \eqref{line_element} and an overhead dot denotes differentiation with respect to proper time.

The MN spacetime, being stationary and axisymmetric, immediately leads to the conservation of the secondary's temporal and azimuthal momenta $p_t,\,p_\phi$, with $p_\lambda=\mu\, g_{\lambda\nu}\dot{x}^\nu$, through the equations of motion $\dot{p}_t=0=\dot{p}_\phi$. There are, therefore, two constant of motion; the energy $E$ and the $z$-component of the angular momentum $L_z$, which are given by
\begin{align}
E&=-\mu\left(g_{tt}\dot{t}+g_{t\phi}\dot{\phi}\right)=\mu \,f\left(\dot{t}-\omega \dot{\phi}\right),\\
L_z&=\mu\left(g_{t\phi}\dot{t}+g_{\phi\phi}\dot{\phi}\right)=\mu\left(f \omega \left(\dot{t}-\omega \dot{\phi}\right)+\frac{\rho^2}{f}\dot{\phi}\right).
\end{align}
Together with the preservation of the secondary's rest mass $\mu$, which results from the conservation of $4$-velocity along a geodesic $g_{\lambda\nu}\dot{x}^\lambda\dot{x}^\nu=-1$, one obtains three constants of motion. Unfortunately, these integrals do not suffice to decouple the secondary's motion in the $(\rho,z)$-plane, in contrast to geodesics in Kerr ($q=0$) which possess a fourth integral of motion, the Carter constant, and the equations of motion in the $(\rho,z)$-plane are separable. Consequently, to obtain the motion in MN spacetime, one requires use of the coupled second-order differential equations \eqref{geodesic}:
\begin{align}
2\,g_{\rho\rho}\,\ddot{\rho}+2\,\dot{\rho}\,\dot{z}\,\partial_z g_{\rho\rho}+\dot{\rho}^2\,\partial_\rho g_{\rho\rho}-\dot{t}^2\,\partial_\rho g_{tt}-2\,\dot{t}\,\dot{\phi}\,\partial_\rho g_{t\phi}\nonumber\\
\label{eqrho}
-\dot{z}^2\,\partial_\rho g_{zz}-\dot{\phi}^2\,\partial_\rho g_{\phi\phi}=0,\\
2\,g_{zz}\,\ddot{z}+2\,\dot{\rho}\,\dot{z}\,\partial_\rho g_{zz}-\dot{\rho}^2\,\partial_z g_{\rho\rho}-\dot{t}^2\,\partial_z g_{tt}-2\,\dot{t}\,\dot{\phi}\,\partial_z g_{t\phi}\nonumber\\
\label{eqz}
+\dot{z}^2\,\partial_z g_{zz}-\dot{\phi}^2\,\partial_z g_{\phi\phi}=0,
\end{align}
where
\begin{align}
g_{tt}=-f,\,g_{t\phi}=\omega f,\,g_{\rho\rho}=g_{zz}=\frac{e^{2\gamma}}{f},\,g_{\phi\phi}=\frac{\rho^2}{f}-\omega^2 f,
\end{align}
and the $t$, $\phi$-momenta are given by
\begin{align}
\label{constants1}
\dot{t}&=\frac{E g_{\phi\phi}+L_z g_{t\phi}}{\mu(g_{t\phi}^2-g_{tt}g_{\phi\phi})}=\frac{\omega f}{\mu\,\rho^2}\left[{L_z} + {\omega E}
\left(\frac{\rho^2}{\omega^2 f^2} -1 \right) \right],\\
\label{constants2}
\dot{\phi}&=-\frac{E g_{t\phi}+L_z g_{tt}}{\mu(g_{t\phi}^2-g_{tt}g_{\phi\phi})}=\frac{f}{\mu\,\rho^2} (L_z-\omega E).
\end{align}
Eqs. \eqref{eqrho} and \eqref{eqz} are sufficient to fully describe the geodesic motion of the secondary in the MN spacetime, while the evolution of the $t,\,\phi$ coordinates can be obtained by integrating \eqref{constants1} and \eqref{constants2}. The combination of \eqref{constants1}, \eqref{constants2} and the conservation of $4$-velocity leads to the constraint equation \cite{luk}
\begin{equation}
\label{constraint}
\dot{\rho}^2+\dot{z}^2 + V_{\textrm{eff}}(\rho,z)=0,
\end{equation}
with 
\begin{eqnarray}
\label{potential}
V_{\textrm{eff}}(\rho,z)\equiv \frac{e^{-2\gamma}}{\mu^2}
\left[
\mu^2f-E^2+\left(\frac{f}{\rho}(L_z-\omega E) \right)^2
\right]
\end{eqnarray}
a $2$-dimensional effective potential. The motion in the $(\rho,z)$-plane may, therefore, be thought of as bound motion in the effective potential \eqref{potential}. The curve defined by $V_{\text{eff}}=0$ is called the curve of zero velocity (CZV), since whenever a geodesic orbit reaches it the velocity components in the $(\rho,z)$-plane vanish. Equation \eqref{constraint}, not only provides an understanding of the general properties of bound orbits in the MN spacetime, but also can be used as a cross-check to verify the convergence of geodesic evolution. For an exhaustive analysis of the properties of the effective potential as well as its dependence on various parameters, we refer the reader to \cite{Gair:2007kr,luk,Lukes-Gerakopoulos:2014dpa,cont11}.

To numerically integrate the second-order equations \eqref{eqrho} and \eqref{eqz}, we generally need to prescribe appropriate initial conditions of the form $(\rho(0),\dot{\rho}(0),z(0),\dot{z}(0))$, besides the parameters describing the particulars of the system, which establish some initial position and velocity to the secondary, at $\tau=0$, within the allowed region of bound motion on the $(\rho,z)$-plane defined by the CZV. We reduce the initial conditions to $(\rho(0),\dot{\rho}(0),z(0))$, with the remaining condition $\dot{z}(0)$ being inferred from Eq. \eqref{constraint}. Despite the fact that the equations of motion are written with respect to proper time for simplicity, GW detectors operate with respect to the inertial time $t$ at very large distances from the EMRI source. This is merely an inconvenience since one can recast \eqref{eqrho}, \eqref{eqz} and \eqref{constraint} to evolve with respect to $t$ by using \eqref{constants1} and the chain rule, e.g. $\dot{\rho}=d\rho/d\tau=\dot{t}\,d\rho/dt$.

\section{Long timescale EMRI modeling: radiation reaction and the kludge scheme}\label{sect:diss}
GR predicts that astrophysically-captured compact objects emit radiation in accordance with the Einstein equations. In the presence of gravitational radiation, the integrals of motion cannot be considered constant anymore, but rather evolve as energy and angular momentum is gradually carried away by GWs. Owing to their small mass ratio, EMRIs radiate adiabatically during very long timescales. As discussed above, on short timescales the orbit is approximately undergoing geodesic motion. By taking into account the structure of the secondary, which induces a gravitational self-force, one can calculate gravitational perturbations at the primary's horizon and infinity to obtain the gravitational waveform and the corresponding fluxes of energy and angular momentum. The gravitational self-force, then, leads to some evolution equations for $E$ and $L_z$ and forms the basis for the adiabatic approximation introduced in \cite{mino,gal}, which is valid only when the change in the momenta is sufficiently small over a single orbit. Over short timescales, the system follows an orbit that neglects radiative backreaction, while over long timescales, the inspiral behaves as a `flow' of the system through a sequence of successively damped geodesics.

Although these fluxes can be completely understood through full-blown self-force calculations, a description of gravitational self-force, up to first-order terms, has been achieved only recently for Kerr \cite{Barack:2018yvs,vandeMeent:2017bcc}. To extend these results to bumpy BHs remains a rather difficult and demanding task. Therefore, to simulate the dissipative effect of $E$ and $L_z$ we employ the hybrid kludge scheme, introduced in \cite{kludge1,gair06}, which mixes a notion of short-timescale motion with an approximate description of the long-timescale radiation effects. This scheme is based on combining exact relativistic expressions for the evolution of the orbital elements with approximate post-Newtonian (PN) formulae for energy and angular momentum fluxes.  Hence, radiation reaction is implicitly included by the particulars of the PN coefficients, and all we need to do is solve the ``instantaneous'' geodesic equations \eqref{geodesic} augmented with the appropriate PN formulae.

The kludge approximation has been thoroughly tested, is widely utilized for the study of EMRIs and has been proven to be remarkably faithful, thus provides a trustworthy mechanism to generate trajectories of generic EMRIs particularly quickly \cite{gair06}.

Here, since we are dealing with a spacetime with an anomalous quadrupole moment (see Eq. \eqref{moments}), we augment the fluxes with an additional PN term which represents the effect of $q$ on the evolution of $E$ and $L_z$ \cite{Barack:2006pq,Gair:2007kr,luk}, through $M_2$. Up to second PN order, the kludge equations for the evolution of an inspiral read \cite{gair06}

\begin{widetext}\begin{align} \label{Edot}\nonumber
\bigg< \frac{dE}{dt}\bigg> &= -\frac{32}{5} \frac{\mu^2}{M^2} \left(\frac{M}{p}\right)^5 (1-e^2)^{3/2}\Bigg[ g_1(e)-\chi\left(\frac{M}{p}\right)^{3/2} g_2(e)\cos\iota -\left(\frac{M}{p}\right) g_3(e)+\pi\left(\frac{M}{p}\right)^{3/2} g_4(e) \\
&-\left(\frac{M}{p}\right)^2 g_5(e) -(M_{2}/M^3) \left(\frac{M}{p}\right)^2 g_6(e)+\frac{527}{96} (M_{2}/M^3) \left( \frac{M}{p}\right)^2 \sin^2\iota
 \Bigg] ,\\\nonumber
\label{Lzdot}
\bigg< \frac{dL_z}{dt}\bigg> &= -\frac{32}{5} \frac{\mu^2}{M} \left(\frac{M}{p}\right)^{7/2}(1-e^2)^{3/2}\Bigg[ g_9(e)\cos\iota+\chi\left(\frac{M}{p}\right)^{3/2} \{g_{10}^{a}(e)-\cos^2\iota g_{10}^{b}(e) \}-\left(\frac{M}{p}\right) g_{11}(e) \cos\iota  \\
&+ \pi\left(\frac{M}{p}\right)^{3/2} g_{12}(e)\cos\iota 
-\left(\frac{M}{p}\right)^2 g_{13}(e) \cos\iota - (M_{2}/M^3) \left(\frac{M}{p}\right)^2 \cos\iota \left(g_{14}(e) - \frac{45}{8}\,\sin^2\iota\right) \Bigg] ,\\\nonumber\\ 
\label{gterms}
g_1(e) &= 1 + \frac{73}{24} e^2  + \frac{37}{96}e^4, \quad\quad\quad\quad\quad\quad \quad\quad\quad\quad\quad\quad 
g_2(e)= \frac{73}{12} + \frac{823}{24} e^2 + \frac{949}{32}e^4
+ \frac{491}{192}e^6 ,\  
\nonumber \\
g_3(e) &= \frac{1247}{336} + \frac{9181}{672} e^2 ,\quad\quad\quad\quad\quad\quad \quad\quad\quad\quad\quad\quad\,\,\,\,
g_4(e)  = 4 + \frac{1375}{48} e^2  ,\
\nonumber \\
g_5(e) &= \frac{44711}{9072} + \frac{172157}{2592} e^2 ,
\quad\quad\quad\quad\quad\quad \quad\quad\quad\quad\quad\,
g_6(e) = \frac{33}{16} + \frac{359}{32} e^2  ,\
\nonumber \\
g_9(e) &= 1 + \frac{7}{8} e^2    ,
\quad\quad\quad\quad\quad\quad\quad\quad\quad\quad\quad\quad\quad\quad\quad\,\,\,
g_{10}(e) = \frac{61}{12} + \frac{119}{8} e^2 +\frac{183}{32}e^4  ,\ 
\\
g_{11}(e) &= \frac{1247}{336} + \frac{425}{336} e^2  ,\quad\quad\quad\quad\quad\quad \quad\quad\quad\quad\quad\quad\,\,\,\,\,
g_{12}(e) = 4 + \frac{97}{8} e^2 ,\
\nonumber \\
g_{13}(e) &= \frac{44711}{9072} + \frac{302893}{6048} e^2 ,\quad\quad\quad\quad\quad\quad \quad\quad\quad\quad\,\,\,\,\, \nonumber
g_{14}(e) = \frac{33}{16} + \frac{95}{16} e^2 ,\ \nonumber\\
g_{10}^{a}(e)&=\frac{61}{24} + \frac{63}{8}e^2   + \frac{95}{64}e^4 ,\ \nonumber
\quad\quad\quad\quad\quad\quad \quad\quad\quad\quad\,\,\,\, \nonumber
g_{10}^{b}(e)= \frac{61}{8} + \frac{91}{4}e^2  + \frac{461}{64}e^4 ,\ \nonumber
\end{align}\end{widetext}
where the orbital elements $e,p,\iota$ (where $p$ is \emph{not} to be confused with the momentum), corresponding to the eccentricity, semi-latus rectum and inclination, respectively, are given by
\begin{equation}\label{orbital_elements} 
e=\frac{r_+ -r_-}{r_+ + r_-},\quad\quad p=\frac{2 r_+ r_-}{r_+ + r_-},\quad\quad \iota=\frac{\theta_+ -\theta_-}{2},
\end{equation}
where $r_{+}(r_{-})$ and $\theta_{+}(\theta_-)$ denote the periapsis (apoapsis) and maximum (minimum) angular values, along the orbit, respectively. At this point we need to admit the following: Eqs. \eqref{Edot} and \eqref{Lzdot} are only valid for Kerr BHs. The augmentation of those fluxes with additional terms to account for the anomalous quadrupoles of MN spacetime \cite{Barack:2006pq,Gair:2007kr,apo09,luk}, which is also a solution to GR, serves as a model to understand the qualitative behavior of orbits but does not capture the exact dynamics of the EMRI. Furthermore, we note that we do not attempt to evolve the other components of the angular momentum or some generalization of the Carter constant in MN spacetime, but only introduce dissipation to the energy and azimuthal angular momentum as in \cite{Gair:2007kr,apo09,luk,Destounis:2021mqv,Lukes-Gerakopoulos:2021ybx}. Even so, since we will only evolve the EMRIs for small intervals of time compared to the radiation reaction timescale, and since we are interested only in particular phenomenological imprints, we do not expect that the inclusion of the aforementioned effects will dramatically obscure the following results.

To construct the inspiral we numerically integrate Eqs. \eqref{eqrho} and \eqref{eqz}, utilizing \eqref{constants1} and \eqref{constants2}, augmented with the PN fluxes \eqref{Edot} and \eqref{Lzdot} and assume linear variations for the energy and angular momentum as in \cite{Canizares}
\begin{align}
\label{update_E}
E_1=\frac{E_0}{\mu}+\bigg< \frac{dE}{dt}\bigg>\bigg|_0 N_r\, T_r,\\
\label{update_L}
L_{z,1}=\frac{L_{z,0}}{\mu}+ \bigg< \frac{dL_z}{dt}\bigg>\bigg|_0 N_r\, T_r,
\end{align}
where $E_0,\,L_{z,0}$ are the initial energy and angular momentum, respectively, $\langle{dE}/{dt}\rangle|_0,\,\langle{dLz}/{dt}\rangle|_0$ are the fluxes calculated at the beginning of the inspiral and $T_r$ is the time that the orbit takes to travel from the periapsis to apoapsis and back. Eqs. \eqref{orbital_elements}, and thus the fluxes \eqref{update_E} and \eqref{update_L}, are updated every $N_r$ cycles for the whole EMRI evolution. To further guarantee numerical accuracy when radiation reaction is included, we calculate the $4$-velocity in each update of the fluxes and check its conservation along a geodesic evolution with initial conditions the energy, angular momentum, position and velocity at every update time step. For all simulations presented herein, we find that they are satisfied to within a part in $\sim 10^9$ for the first $10^4$ crossings through the equatorial plane $z=0$.

\section{Imprints of non-integrability}\label{orbital level}
 
Due to stationarity and axisymmetry, the Kerr metric admits two integrals of motion which respectively imply the conservation of energy and angular momentum of relativistic particles. Together with the preservation of the particle's rest mass and the Carter constant, geodesic motion in Kerr is integrable and the orbits are bound in $2$-dimensional tori of the $4$-dimensional phase space. In general, orbits oscillate between the two available degrees of freedom with characteristic frequencies $\omega_1$ and $\omega_2$. The ratio of these frequencies, $\nu_\theta=\omega_1/\omega_2$, is called the rotation number, and carries pertinent information about the features of the orbit. If the ratio of the frequencies is an irrational number, the corresponding orbit winds around its torus, covering it densely. Such orbits are called quasiperiodic. In special cases, the ratio of frequencies becomes a rational number $n/m$, where $n,\,m \in \mathbb{N}$, and the orbit returns to its starting point after $m$ windings. These orbits are called periodic and the corresponding torus is called resonant. The phase-space trajectory of a periodic orbit is not space-filling, in contrast to quasiperiodic orbits.

When monitoring the trajectory's successive intersections on a particular $2$-dimensional surface of section within the phase space (Poincar\'e map), one can define closed curves for each torus, named invariant curves \cite{Lichtenberg92}. A quasiperiodic orbit intersects the surface of section repeatedly and covers densely the corresponding invariant curve which surrounds a fixed central point of the map. On the other hand, the intersections of periodic orbits form resonant invariant curves consisting of an infinite number of $m$-multiplets of periodic points. When the system in study is non-integrable, the resonant invariant curves nest around periodic stable points, thus forming islands of stability (Birkhoff islands), in accordance with the Kolmogorov-Arnold-Moser (KAM) \cite{moser62,arnold63} and Poincar\'e-Birkhoff theorems\cite{poin12,birk13}. 
 
By tracking the angles $\vartheta$ formed between subsequent intersections on a surface of section, relative to some fixed point, we can define a sequence of rotation numbers, 
\begin{equation} 
\label{rotation}
\nu_{\theta,N} = \frac {1} {2 \pi N} \sum^{N}_{i=1} \vartheta_{i},
\end{equation}
where in the limit $N \rightarrow \infty$, the sequence \eqref{rotation} converges to $\nu_\theta$. Generally, integrable systems exhibit a monotonous change in consecutive rotation numbers (rotation curve), while non-integrable systems exhibit plateaus \cite{cont02,luk,dest20} when the orbit is crossing a Birkhoff island.
\begin{figure*}[t]
	\includegraphics[scale=0.34]{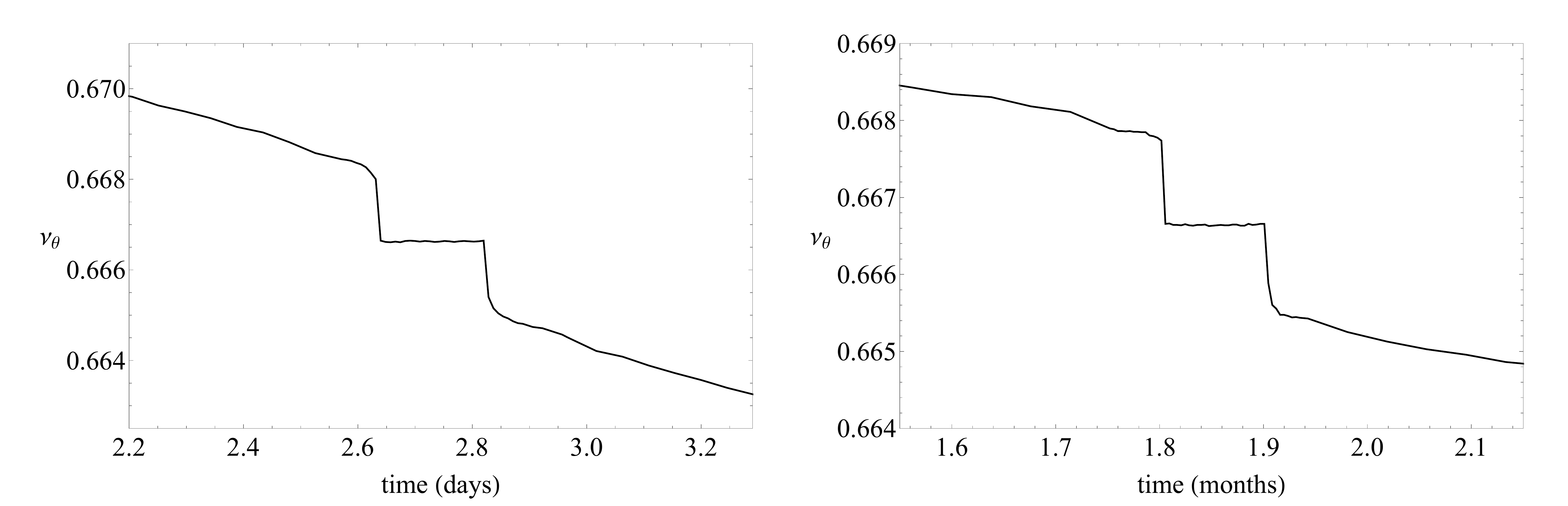}
	\caption{Dissipative rotation curves for the EMRI cases $\text{(i)}$ (left) and $\text{(i)}$ (right). Both plateaus appear at $\nu_\theta=2/3$, designating the crossing of the $2/3$-resonant island.}
	\label{rot_curve}
\end{figure*}
The rotation curve has been utilized to distinguish whether various EMRI systems have integrable equations of motion (see \cite{apo09,luk,LukesGerakopoulos:2012pq,Lukes-Gerakopoulos:2014dpa,cont11,Zelenka:2017aqn,Zelenka:2019nyp,dest20}). At the geodesic level, to form a rotation curve, one needs to calculate successive rotation numbers, with respect to a fixed point $\boldsymbol{u}_0$, by smoothly increasing one of the phase space parameters, e.g. the radius, while keeping the rest fixed. As explained above, the calculation of rotation numbers involves the geodesic evolution of an orbit whilst keeping track of all crossings through a particular surface of section. The longer we evolve the geodesic, the more intersection points we obtain, with the number of intersections being proportional to the precision of the rotation number (e.g. $\sim 10^4$ crossings gives $4$ digits of precision to the resulting rotation number). A typical graph of a rotation curve in MN spacetime can be found in \cite{luk} (see Figure 3 therein).

Rotation curves are particularly helpful to determine the existence of non-integrability in EMRIs even when radiation reaction is included. After evolving an inspiral to obtain the trajectory of the secondary around a primary MN compact object, one can analyze the $\rho$ and $z$ time-series in the frequency domain with Fourier analysis, and infer the rotation numbers in successive time segments \cite{luk}. This leads to a rotation curve which evolves with time, and exhibits a plateau during the passage through a resonant island. An alternative way of producing the rotation curve of an inspiral, which we utilize here, was recently proposed in \cite{Lukes-Gerakopoulos:2021ybx}. The main idea is to take various points from the dissipative evolution of $\rho$, $z$, $E$ and $L_z$ as initial conditions and evolve those particular points as geodesics for enough time in order to obtain the respective rotation number of each point. 

In what follows, we consider two inspirals, with the following parameters and initial conditions, so that they pass through the $2/3$-resonance
\begin{enumerate}
	\item  $\mu/M=10^{-5},\, [\rho(0),\dot{\rho}(0),z(0)]=(4.23M,0,0)$,
	\item  $\mu/M=10^{-6},\, [\rho(0),\dot{\rho}(0),z(0)]=(4.129M,0,0)$,
\end{enumerate}
where we fix the parameters $E_0=0.95\mu$, $L_{z,0}=3M\mu$, $\chi=0.9$ and consider a light secondary object of $\mu=1M_\odot$ inspiraling into a primary MN supermassive compact object with deformation parameter $q=0.95$, where its mass $M$ is defined by the respective mass ratio ($M=10^5 M_\odot$ for $\text{(i)}$ and $M=10^6 M_\odot$ for $\text{(ii)}$). As detailed in Sec. \ref{sect:geod}, the remaining initial condition $\dot{z}(0)$ is fixed by the constraint equation \eqref{constraint}. The total evolution time of the EMRIs are $t_\text{evol}=2\times 10^6 M\simeq 3.5$ months for case $(\text{i})$ and $t_\text{evol}=10^6 M\simeq5.5$ days for case $(\text{ii})$. Both cases evolve with an approximate orbital period of $T_r\sim 2\times 10^2M$, which translates to $\sim 5\times 10^3$ and $\sim 10^4$ total cycles, for EMRI $(\text{i})$ and $(\text{ii})$, respectively.

In Fig. \ref{rot_curve} we present the dissipative rotation curves for the cases $(\text{i})$ and $(\text{ii})$ discussed above. A clear plateau appears in both EMRIs, at roughly half of their respective $t_\text{evol}$. The plateaus are fixed at $\nu_\theta=2/3$, which designates the time it takes for the secondary to cross the $2/3$-resonant island. Occupancy in the island lasts for $\sim 200$ cycles, in both cases, and should lead to a rather significant effect on the characteristics of the orbit.

\begin{figure*}[t]
	\includegraphics[scale=0.34]{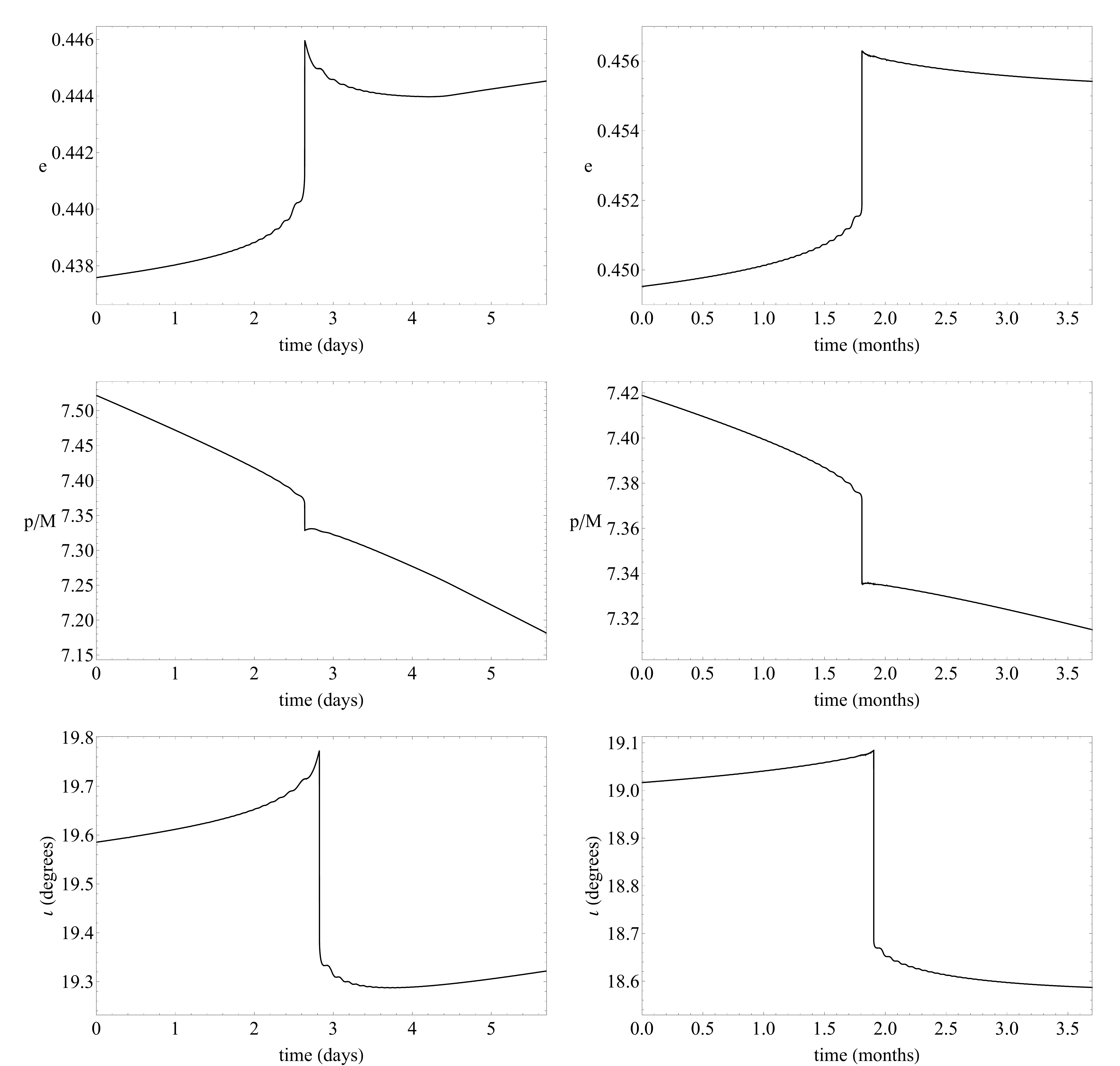}\hskip 3ex
	\caption{Eccentricity (top row), semi-latus rectum (middle row) and inclination (bottom row) evolution for the EMRI cases $\text{(i)}$ (left column) and $\text{(i)}$ (right column). The abrupt changes occur during the $2/3$-resonant island crossing for both cases.}
	\label{orbital}
\end{figure*}

The evolution of the orbital elements during the inspiral is shown in Fig. \ref{orbital}. Indeed, a very abrupt change occurs, during passage through the resonant island, with the semi-latus rectum, eccentricity and inclination undergoing a sudden shift of $\sim0.6\%$, $1\%$ and $2\%$, respectively. These `kicks' are quite sudden and occur at entry into the resonant island. Note that away from resonant islands, we update the fluxes every $10$ cycles, i.e. $N_r=10$. Close to and during passage through an island, we increase the update rate to $1$ per cycle to properly capture such dynamical phenomenon. This leads to $\sim 250$ updates around the vicinity of a resonance. We further note that for the calculation of the orbital elements (and the subsequent fluxes per update) we evolved geodesics with appropriate initial conditions for $2.5\times 10^3$ and $10^4$ cycles and obtained convergence of up to $7$ digits. Since the sudden jumps in the orbital elements occur on around the third digit, we are reassured that the behavior portrayed in Fig. \ref{orbital} is qualitatively valid.
We expect that the corresponding kicks resulting from a hypothetical instantaneous self-force evolution, which includes also conservative terms affecting the immediate motion, would be smoother, in accordance with the findings in \cite{Flanagan:2010cd,Berry:2016bit}. Even so, the fact that we can observe any effect in the orbital elements during passage through a resonant island with the adiabatic approximation is quite fascinating, although such approach is inadequate in capturing the full extent of this phenomenon.

In the following sections, we will demonstrate that the abrupt shifts in the orbital elements, caused during passage through a resonant island, is directly imprinted in the gravitational waveform of a non-integrable EMRI.

\section{EMRI Waveform modeling and detector response}

Although crossing a resonant island serves as a clear signature of non-integrability in EMRIs, it is quite challenging for one to extract the fundamental orbital frequencies from a waveform in order to sketch a rotation curve. GWs, indeed, carry information regarding the inspiral's fundamental orbital frequencies, though in most cases these quantities are hidden within the waveform as linear combinations of all orbital constituents. In this section, we investigate the GW emission of the aforementioned demonstrative EMRIs. We will show that the frequency evolution of an incoming GW, detected by LISA, contains a clear imprint of the passage through a resonant island, in the form of an abrupt, but consistent, frequency `glitch', in a similar manner as those observed in \cite{Destounis:2021mqv}.

To model the gravitational waveform of island-crossing EMRIs we adopt the `quick and dirty' numerical kludge scheme, which combines exact particle trajectories with approximate GW emission \cite{glamp07}.  A more sophisticated approach could involve some (appropriately generalized, see e.g. \cite{suv19a,suv19b}) Teukolsky equations to deduce directly the GW characteristics through the Weyl scalars. However, such an analysis is considerably more complicated and does not help to elucidate the main new features presented in this work: the universality of `glitches' in non-integrable EMRIs. The numerical kludge is perfectly-suited for phenomenology and has demonstrated remarkable agreement ($\sim 95\%$) with Teukolsky-based waveforms \cite{glamp07}.

The radiative component of a metric perturbation at large luminosity distance $d$ from a source $\boldsymbol{T}$, within the Einstein-quadrupole approximation, can be written, in the transverse-traceless (TT) gauge, as \cite{thorne80,glamp07,Canizares}
\begin{equation} \label{eq:metpert}
	h^{\text{TT}}_{ij}=\frac{2}{d} \frac {d^2 I_{ij}} {dt^2}
\end{equation}
where $I_{ij}$ is the symmetric and trace-free (STF) mass quadrupole associated to the perturbation,
\begin{equation}
	I^{ij}=\left[\int d^3x \,x^i x^j\, T^{tt}(t,x^i)\right]^\text{STF},
\end{equation}
with $t$ the inertial time measured by the detector. The $tt-$component of the stress-energy tensor for a point-particle with trajectory $\boldsymbol{Z}(t)$ reads \cite{peters69}
\begin{equation}\label{Ttt}
	T^{tt}(t,x^i) = \mu\delta^{(3)} \left[x^i - Z^i(t) \right].
\end{equation}
where $\delta^{(3)}$ denotes the $3$-dimensional Dirac delta distribution. Utilizing Eqs. \eqref{BoyLindr} and \eqref{BoyLindtheta}, one can transform the quasi-cylindrical coordinates used in Eq. \eqref{line_element}, $(\rho,z)$, to the standard Boyer-Lindquist coordinates $(r,\theta)$. These, asymptotically reduce to spherical coordinates, so that suitable pseudo-flat coordinates (not to be confused with the prolate spheroidal coordinates $(x,y)$)
\begin{align} \label{eq:cartesian}
	x=r \sin\theta \cos\phi,\,\,\,\,\,\, y=r \sin\theta \sin\phi,\,\,\,\,\,\,z=r \cos\theta,
\end{align} 
can be identified, and related to the position of a space-borne detector at infinity. In reality, the detector is located at some finite $d$, so this prescription is not strictly speaking valid, though has been shown to model EMRI waveforms quite well when compared to those produced using more sophisticated approaches \cite{glamp07,Sopuerta:2011te}.
\begin{figure}[t]
	\includegraphics[scale=0.265]{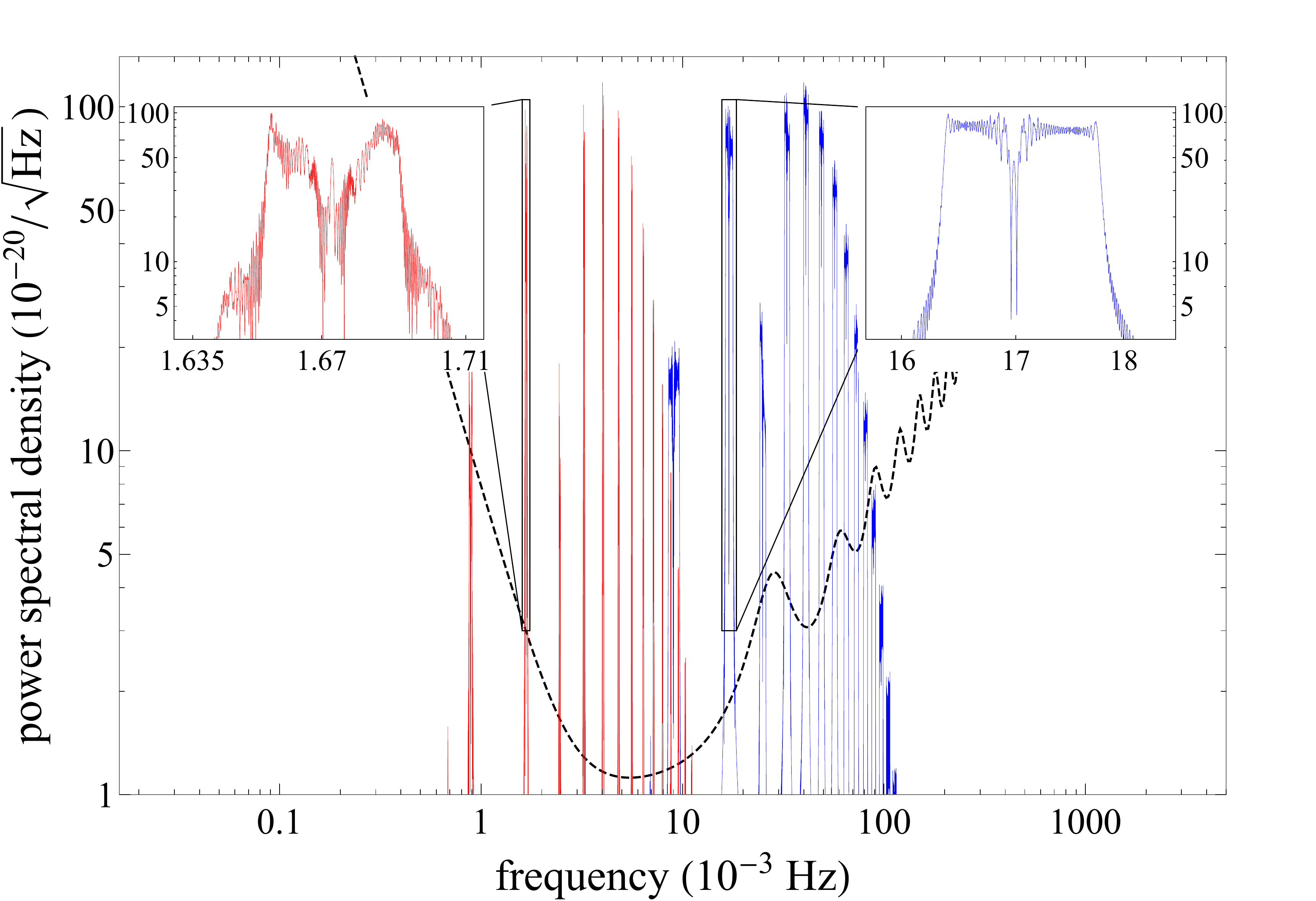}
	\caption{Power spectral density of case $\text{(i)}$ (blue) and $\text{(ii)}$ (red) EMRIs at a luminosity distance $d=50\,\text{Mpc}$, over-plotted on the LISA power spectral density sensitivity curve (dashed black curve). The most prominent Fourier peaks are presented in the zoomed inlays, while the remaining peaks correspond to higher harmonics.}
	\label{detectability}
\end{figure}
An incoming GW can be projected onto its mutually orthogonal $+$ and $\times$ polarization states by introducing two vectors $\boldsymbol{p} = \boldsymbol{n} \times {\boldsymbol{z}} / |\boldsymbol{n} \times {\boldsymbol{z}}|$ and $\boldsymbol{q} = \boldsymbol{p} \times \boldsymbol{n}$, which are defined in terms of a unit vector $\boldsymbol{n}$ which points from the source to the direction of the detector and the unit vector $\boldsymbol{z}$ designating the spin direction, $\boldsymbol{S}=a M \boldsymbol{z}$. In terms of the polarization tensors
\begin{equation}
	\epsilon_+^{ij}=p^i p^j-q^i q^j,\,\,\,\,\,\,\epsilon_\times^{ij}=p^i q^j+p^j q^i,
\end{equation}
constructed by the spatial orthonormal basis $(\boldsymbol{n}, \boldsymbol{p},\boldsymbol{q})$, the corresponding GW metric perturbation is
\begin{equation}
	h^{ij}(t)=\epsilon_+^{ij}h_+(t)+\epsilon_\times^{ij}h_\times(t),
\end{equation}
with
\begin{equation}
	h_+(t)=\frac{1}{2}\epsilon_+^{ij}h_{ij}(t),\,\,\,\,\,\,\,h_\times(t)=\frac{1}{2}\epsilon_\times^{ij}h_{ij}(t).
\end{equation}
To express the GW components in terms of the position, $Z^i(t)$, velocity, $v^i(t)=dZ^i/dt$, and acceleration vectors $a^i(t)=d^2Z^i/dt^2$, we use Eqs. \eqref{eq:metpert} and \eqref{Ttt} to obtain \cite{Canizares}
\begin{equation}
	\label{GW_formula}
	h_{+,\times}(t)=\frac{2\mu}{d}\epsilon^{+,\times}_{ij}\left[a^i(t)Z^j(t)+v^i(t)v^j(t)\right].
\end{equation}
In general, LISA's response to an incident GW is determined by a vector $h_{\alpha}$ which depends on the antenna pattern response functions $F^{+,\times}_{I,II}$ and the  $+$ and $\times$ polarization states through \cite{barack04}
\begin{equation}
	h_{\alpha}(t)=\frac{\sqrt{3}}{2}\left[F^{+}_{\alpha}(t)h_{+}(t)+F^{\times}_{\alpha}(t)h_{\times}(t)\right],
\end{equation}
where $\alpha=(I,II)$ is an index representing the different antenna pattern functions \cite{Apostolatos:1994mx,Cutler:1997ta,barack04,Moore:2017lxy}
\begin{align}
	F^{+}_I &= \frac{1}{2}(1+\cos^2\Theta)\cos(2\Phi)\cos(2\Psi) \nonumber \\
	&-\cos\Theta\sin(2\Phi)\sin(2\Psi),\label{responseIp}\\
	F^{\times}_I &= \frac{1}{2}(1+\cos^2\Theta)\cos(2\Phi)\cos(2\Psi)\nonumber \\
	&+\cos\Theta\sin(2\Phi)\sin(2\Psi),\label{responseIc}\\
	F^{+}_{II} &= \frac{1}{2}(1+\cos^2\Theta)\sin(2\Phi)\cos(2\Psi) \nonumber 
	\\&+\cos\Theta\cos(2\Phi)\sin(2\Psi), \label{responseIIp}\\
	F^{\times}_{II} &= \frac{1}{2}(1+\cos^2\Theta)\sin(2\Phi)\sin(2\Psi) \nonumber \\
	&-\cos\Theta\cos(2\Phi)\cos(2\Psi), \label{responseIIc}
\end{align}
where $(\Theta,\Phi)$ is the sky location and $\Psi$ the polarization of the source in a precessing detector-based coordinate system. Assuming a fixed orientation $\boldsymbol{n} = (0,0,1)$ and that the supermassive compact object's spin polar and spin azimuthal angles remain fixed at the equatorial plane for simplicity, the angles $(\Theta,\Phi,\Psi)$ introduced in \eqref{responseIp}-\eqref{responseIIc} read (see Ref. \cite{Canizares} for formulae in the general case) $\Theta(t) = \pi/3$, $\Phi(t) = 2 \pi t /T + \pi/2$, and $\Psi = - 2 \pi t /T$, where $T=1$ year is the orbital period of the Earth around the Sun. Although the LISA data stream consists of two linearly independent channels, we will omit one channel for simplicity and neglect any noise in the data stream to demonstrate the main characteristics of detection, therefore, we assume a data stream $s(t)=h_\alpha(t)$.

\section{Detectability and Frequency evolution}
\begin{figure*}[t]
	\includegraphics[scale=0.3]{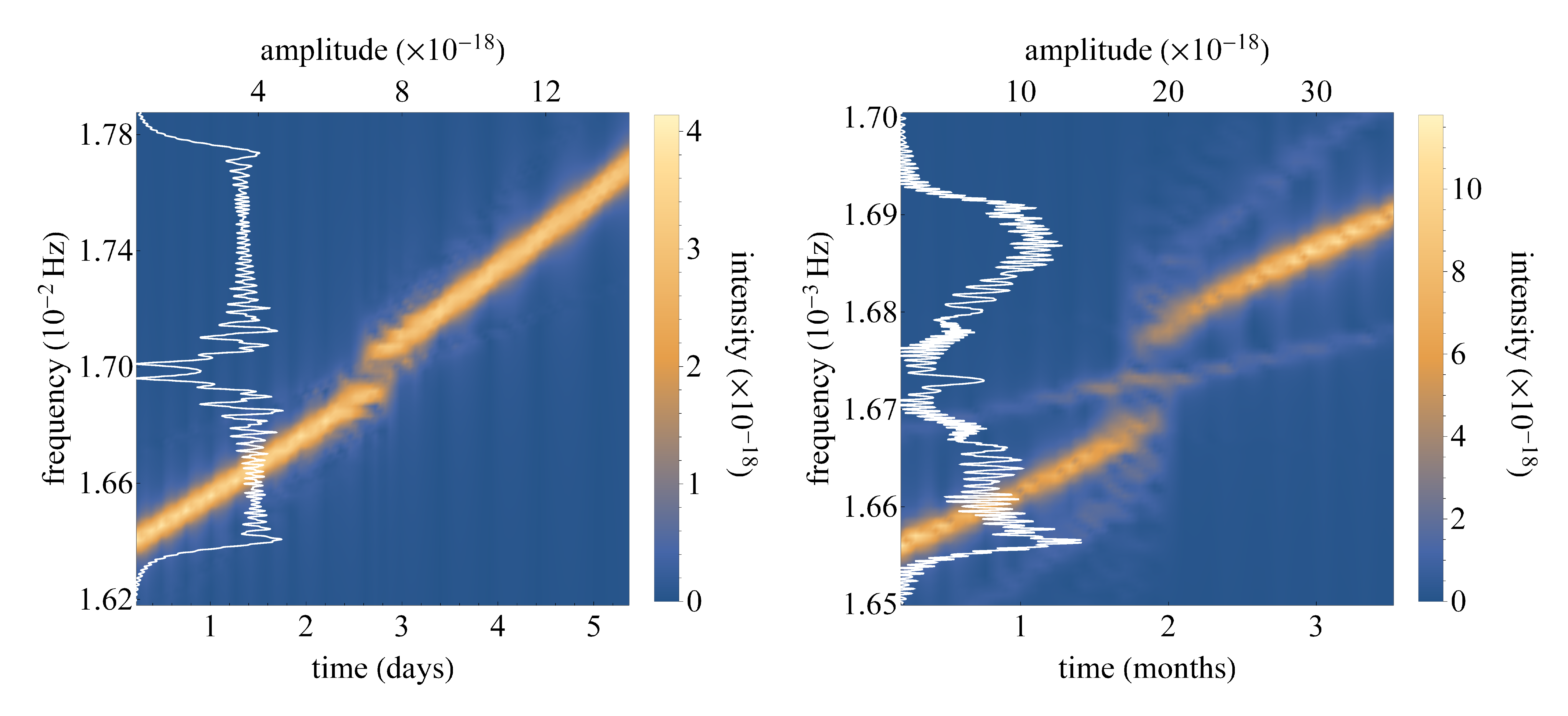}
	\caption{Frequency evolution diagrams for the EMRI cases $\text{(i)}$ and $\text{(ii)}$ depicted in the inlays of Fig. \ref{detectability}.}
	\label{glitches}
\end{figure*}
Here, we consider the EMRI sources $(\text{i})$ and $(\text{ii})$ from Sec. \ref{orbital level} and assume a luminosity distance $d=50\, \text{Mpc}$ from the LISA detector. To test the detectability of GWs from these sources we utilize the square root of the power spectral density (PSD). When discussing a detector, we can define the square of the PSD $S_h(f)$ for the source as \cite{Moore:2014lga}
\begin{equation}
\sqrt{S_h(f)}=2 \sqrt{f}|\tilde{h}(f)|,
\end{equation}
where $f$ the frequency range and $|\tilde{h}(f)|$ the magnitude of the Fourier transform of the signal. The root PSD is the most frequently plotted quantity in the literature and has units of $\text{Hz}^{-1/2}$.

Fig. \ref{detectability} displays the root PSD of EMRIs $(\text{i})$ and $(\text{ii})$ on top of a typical LISA sensitivity curve. A plethora of peaks are present in the spectrum which correspond to the fundamental frequencies and higher harmonics of the GW signals. As expected, such sources produce clearly observable signals, in the vicinity of LISA's maximum sensitivity ($\sim 10^{-3}-10^{-2}$ Hz). Although a Kerr (or, in general, any integrable) EMRI produces peaks which smoothly vary over the available frequency range (see Figure 1 in \cite{Destounis:2021mqv}), non-integrable EMRIs have richer Fourier structures and form amplitude valleys whilst crossing the $2/3$-resonant island, where the PSD peaks drop by up to $2$ orders of magnitude (see inlays in Fig. \ref{detectability}). Consequently, the formation of valleys in the Fourier peaks of a GW signal designates the passage through an island and the non-integrability of the EMRI equations of motion, in a precise manner.

Despite the fact that the Fourier peak structure of non-integrable EMRIs implies the existence of transient frequency modulation during island crossings, the frequency dynamics are better captured through spectrograms. To construct a spectrogram we use the short-time Fourier transform (STFT); a typical method used to determine the frequency content of local sections of a signal as it changes over time. By dividing the waveform into shorter segments of equal length, we compute the Fourier transform separately on each segment, thus revealing its frequency content. We then plot the change of the frequencies from one temporal segment to another, eventually obtaining the frequency evolution. To achieve a smoother transition between segments, we employ a window of fixed size with an offset, which enables it to slide over the available signal. At each slide the Fourier transform is evaluated. This method allows overlapping of time segments and leads to smoother frequency transitions. One of the pitfalls of the STFT is that it has a fixed resolution. The width of the window is directly related to how one wants to represent the signal. A wide window (wideband transform) gives better frequency resolution though poor time resolution. On the other hand, a narrow window (narrowband transform) gives good time resolution but poor frequency resolution. Therefore, the choice of window size is related to an uncertainty principle as the standard deviation in frequency and time is limited. In what follows, we employ a window with appropriate offset so that we represent the frequency evolution of the given signals as optimal as possible, though paying more attention to frequency resolution.

In Fig. \ref{glitches} we present the evolution of the fundamental frequencies of the EMRI cases $(\text{i})$ and $(\text{ii})$ with respect to time. Both cases exhibit an abrupt, but consistent, glitch when the valley of the peaks is observed. Smaller mass ratios (such as the one in case $(\text{ii})$) tend to display more prominent frequency jumps, as a result of the slower passage of the orbit through successive geodesics. In actuality, even though the glitch in case $(\text{ii})$ seems more distinguishable than that of case $(\text{i})$, both inspirals spend a similar amount of cycles in the island and span over a comparable frequency range. These phenomenon is shared throughout the whole GW spectrum, though higher harmonics tend to display less prominent effects. The results presented here, are coherent with those in \cite{Destounis:2021mqv} and demonstrate that glitches are universal and signify the presence of Birkhoff islands and chaotic phenomena in non-integrable EMRIs, even when the primary is a solution to GR.

\section{Discussion} 
In this work, we have studied EMRI orbits and waveforms of a non-Kerr supermassive compact object with anomalous multipolar structure, with respect to that of Kerr, namely the MN spacetime. By assuming that such deformations from a Kerr description may be present in the EMRI composition, we are left with a non-integrable system of equations of motion. In turn, non-integrability implies the existence of chaos \cite{cont02}. 

Nevertheless, full-blown chaos is not expected to be a prominent effect in EMRIs since the occurrence of ergodic orbits is rather rare. The phase-space structure of non-integrable EMRIs is rather similar to that of an integrable one in the majority of the available phase space \cite{arnold63}. Even so, when the inspiral is close to a resonance, the structure of phase space is fundamentally altered. Islands of stability surround resonances \cite{poin12,birk13}, within which the ratio of the radial and polar orbital frequencies, namely the rotation number, is rational and shared throughout the island. Since resonances are abundant in astrophysically-captured bodies \cite{Brink:2013nna}, the identification of `smoking-gun' phenomena associated with non-integrability will shed light into the nature of EMRIs and the role of fundamental spacetime symmetries. 

We find that at the orbital level, a clear imprint evinces when the test body transverses a resonant island; besides the fact that the rotation number displays a dynamical plateau \cite{apo09,luk,cont11,Lukes-Gerakopoulos:2014dpa,Zelenka:2017aqn,dest20}, the orbital elements undergo rapid changes when the secondary enters the resonant island. We expect that these abrupt manifestations would be smoother when one considers an instantaneous self-force evolution \cite{Flanagan:2010cd,Flanagan:2012kg,Berry:2016bit}. Recently, these plateaus were associated with an abrupt jump in the GW frequency evolution, thus providing a definitive signature of non-integrability at the waveform level \cite{Destounis:2021mqv}.

We have extended such waveform analysis to an exact vacuum solution of the Einstein field equations, such as the MN spacetime \cite{Manko92}. The existence of GW glitches during resonant-island crossings appears to be a generic attribute of non-integrable EMRIs, and not just an artifact of theory-agnostic spacetimes. These abrupt jumps in the GW frequencies are sourced by the rapid changes on the orbital elements, even at the adiabatic-approximation level. Despite the fact that Kerr EMRIs also undergo resonant kicks \cite{Ruangsri:2013hra}, their effect can only be revealed with instantaneous self-force calculations \cite{Berry:2016bit}, and evince themselves in the form of accumulated dephasing \cite{Flanagan:2010cd}. One can extrapolate how strong will be the impact of resonant islands in parameter estimation, with respect to a typical Kerr resonance, since the orbit experiences prolonged resonant motion for hundreds of cycles, without taking into account near-resonance motion. 

Ultimately, the adiabatic approximation of EMRIs around resonances should be replaced by full self-force computations, though such task is highly challenging, especially for bumpy EMRIs with complicated multipolar structure. A first step towards this direction could be the augmentation of the numerical kludge scheme utilized here, with an effective resonance model, such as the one recently proposed in \cite{Speri:2021psr}. Another direction would be to utilize Newtonian analogs, such as the one in \cite{Eleni:2019wav}, to understand the phenomenology of island crossings under instantaneous self-force dissipation, though a Newtonian analog prescription of a non-Kerr spacetime is still lacking.

We have argued that GW glitches are universal in non-Kerr EMRIs and have a direct astrophysical origin. In our particular case, they function as a signature of non-Kerrness, though we expect that similar effects should occur in $N>2$-body EMRIs \cite{Barausse:2006vt,AmaroSeoane:2011id} or when the multipole structure of the small-mass companion is taken into account \cite{kiu04,Zelenka:2019uke,Zelenka:2019nyp}. Interestingly, instrumental glitches have been spotted in LISA pathfinder data \cite{Edwards:2020tlp}, though such events in the acceleration noise are rather abrupt and have a distinct structure with respect to the glitches shown here. Therefore, our results serve as a phonomenological imprint of non-integrability and portray a ``smoking-gun'' of chaotic phenomena in EMRIs.

\section*{Acknowledgements}
The authors are indebted to Arthur G. Suvorov for helpful discussions.

\end{document}